\documentclass[12pt,draftclsnofoot, perreview, onecolumn]{IEEEtran}
\usepackage{amsfonts}
\usepackage{mathrsfs}
\usepackage{amsmath}
\usepackage{amssymb}
\usepackage{enumerate}
\usepackage{color,xcolor}
\usepackage{graphicx}
\usepackage{subfigure}
\usepackage{pifont}

\ifCLASSINFOpdf
\else
\fi

\newtheorem{proposition}{Proposition}
\newtheorem{algorithm}{\textbf{Algorithm}}
\begin{document}
%
\title{Serial Concatenation of RS Codes with Kite Codes: Performance Analysis, Iterative Decoding and Design}
%
%

\author{Xiao~Ma, Kai Zhang, Baoming~Bai and Xiaoyi~Zhang
\thanks{X.~Ma and K.~Zhang are with the Department
of Electronics and Communication Engineering, Sun Yat-sen
University, Guangzhou 510275, China. (E-mail: maxiao@mail.sysu.edu.cn)}
\thanks{B.~Bai is with State Key Lab.~of ISN, Xidian University, Xi'an 710071, China. (E-mail: bmbai@mail.xidian.edu.cn)}
\thanks{X.~Zhang is with National Digital Switching System Engineering and Technological R\&D Center, Zhengzhou 450002, China.}
}

\maketitle

\begin{abstract}
In this paper, we propose a new ensemble of rateless forward error
correction~(FEC) codes. The proposed codes are serially
concatenated codes with Reed-Solomon~(RS) codes as outer codes and
Kite codes as inner codes. The inner Kite codes are a special
class of prefix rateless low-density parity-check~(PRLDPC) codes,
which can generate potentially infinite~(or as many as required)
random-like parity-check bits. The employment of RS codes as outer
codes not only lowers down error-floors but also ensures~(with
high probability) the correctness of successfully decoded
codewords. In addition to the conventional two-stage decoding,
iterative decoding between the inner code and the outer code
are also implemented to improve the performance further. The
performance of the Kite codes under maximum likelihood~(ML)
decoding is analyzed by applying a refined Divsalar bound to the
ensemble weight enumerating functions~(WEF). We propose a
simulation-based optimization method as well as density evolution~(DE)
using Gaussian approximations~(GA) to design the Kite codes.
Numerical results along with semi-analytic bounds show that the
proposed codes can approach Shannon limits with extremely low
error-floors. It is also shown by simulation that the proposed
codes performs well within a wide range of
signal-to-noise-ratios~(SNRs).
\end{abstract}


%
\IEEEpeerreviewmaketitle

\section{Introduction}\label{introduction}
%
%
%
%

Serially concatenated codes were first introduced by
Forney~\cite{Forney66}. A traditional concatenated code typically consists
of an outer Reed-Solomon~(RS) code, an inner convolutional code
and a symbol interleaver in between. A typical example is the
concatenation of an outer $\textrm{RS}[255, 223, 33]$ code and an inner
convolutional code with rate 0.5 and constraint length 7, which
has been adopted as the CCSDS Telemetry
Standard~\cite{Costello98}. The basic idea behind the concatenated
codes is to build long codes from shorter codes with manageable
decoding complexity. The task of the outer RS decoder is to
improve the performance further by correcting ``remaining" errors
after the inner decoding. Since the invention of turbo
codes~\cite{Berrou93} and the rediscovery of low-density
parity-check~(LDPC) codes~\cite{Gallager62}, it has been suggested
to use serially concatenated codes with outer RS codes and inner
turbo/LDPC codes for data
transmissions~\cite{ITU-T04}\cite{ETSI06}.

Binary linear rateless coding is an encoding method that can
generate potentially infinite parity-check bits for any given
fixed-length binary sequence. Fountain codes constitute a class of
rateless codes, which were first mentioned in~\cite{Byers98}. For
a conventional channel code, coding rate is well defined; while for a
Fountain code, coding rate is meaningless since the number of
coded bits is potentially infinite and may be varying for
different applications. The first practical realizations of
Fountain codes were LT-codes invented by Luby~\cite{Luby02}.
LT-codes are linear rateless codes that transform $k$ information
bits into infinite coded bits. Each coded bit is generated
independently and randomly as a binary sum of several randomly selected information bits,
where the randomness is governed by the so-called robust soliton distribution. LT-codes have encoding
complexity of order $\log k$ per information bit on average. The
complexity is further reduced by Shokrollahi using Raptor
codes~\cite{Shokrollahi06}. The basic idea behind Raptor codes is
to precode the information bits prior to the application of an
appropriate LT-code. For relations among random linear Fountain
codes, LT-codes and Raptor codes and their applications over
erasure channels, we refer the reader to~\cite{MacKay05}.

Fountain codes were originally proposed for binary erasure channels~(BECs). In addition to their success in erasure channels,
Fountain codes have also been applied to other noisy binary input memoryless symmetric
channels~(BIMSCs)~\cite{Palanki04}\cite{Etesami06}\cite{Pakzad06}.
Comparing other noisy BIMSCs with BECs, we have the following observations.
\begin{enumerate}
  \item Over erasure channels, Fountain codes can be used in an asynchronous way as follows. The transmission bits are first grouped into packets. Then
        LT-codes and/or Raptor codes are adapted to work in a packet-oriented way which is similar to the bit-oriented way. The only difference lies in that the encoders can insert a {\em packet-head} to each coded packet, specifying the involved information packets. While the receiver receives a coded packet, it also knows the connections of this coded packets to information packets. In this scenario, the receiver does not necessarily know the code structure before the transmission. The overhead due to the insertion of packet-head can be made negligible whenever the size of the packets is large.  However, over other noisy channels, it should be more convenient and economic to consider only synchronous Fountain codes. This is equivalent to require that the receiver knows the structure of the code before transmission or shares common randomness with the transmitter. Otherwise, one must find ways to guarantee with probability 1 the correctness of the packet-head.
  \item Both LT-codes and Raptor codes have been proved to be universally good for erasure channels in the sense that, no matter what the channel erasure probability is, the iterative belief propagation~(BP) algorithm can recover the transmitted $k$ bits~(packets) with high probability from any received $k(1+\epsilon)$ bits~(packets). However, it has been proved~\cite{Etesami06} that no universal Raptor codes exist for BIMSCs other than erasure channels.
  \item Over erasure channels, the receiver knows exactly when it successfully recovers the information bits and hence the right time to stop receiving coded bits. However, over other BIMSCs, no obvious ways~(without feedback) exist to guarantee  with probability 1 the correctness of the recovered information bits.
\end{enumerate}

In this paper, we are concerned with the construction of synchronous rateless codes for binary input additive white Gaussian noise~(BIAWGN) channels. We propose a new ensemble of rateless forward error
correction~(FEC) codes based on serially concatenated structure, which
take RS codes as outer codes and Kite codes as
inner codes. The inner Kite codes are a special class of prefix
rateless low-density parity-check~(PRLDPC) codes, which can
generate as many as required random-like parity-check bits. The
use of RS codes as outer codes not only lowers down error-floors
but also ensures~(with high probability) the correctness of
successfully decoded codewords. The proposed codes may find
applications in the following scenarios.
\begin{enumerate}

    \item The proposed encoding method can definitely be utilized
    to construct FEC codes with any given coding rate. This
    feature is attractive for future deep-space communications~\cite{Calzolari07}.

    \item The proposed codes can be considered as rate-compatible
    codes, which can be applied to hybrid automatic repeat
    request~(H-ARQ) systems~\cite{Hagenauer88}.

    \item The proposed codes consist of a family of FEC codes
    with rates varying ``continuously" in a wide range.
    When combined with bit-interleaved coded modulation~(BICM)~\cite{Caire98},
    the proposed coding can be applied to adaptive coded modulation~(ACM)
    systems~\cite{Goldsmith05}.

    \item The proposed codes can be utilized to broadcast common
    information over unknown noisy channels~\cite{Shulman04}.
    For example, assume that a sender is to transmit simultaneously a common message using binary-phase-shift-keying~(BPSK)
    signalling to multiple receivers through different additive white Gaussian noise~(AWGN) channels. Especially when these
    channels are not known to the sender, for the same reason that Fountain codes were motivated for erasure channels, rateless
    FEC codes can provide an effective solution.
\end{enumerate}

{\bf Structure:} The rest of this paper is organized as follows. In
Section~\ref{sec2}, we present the constructions of the Kite codes
as well as their encoding/decoding algorithms. The relationships between the Kite codes and existing codes are discussed. We emphasized that the Kite codes are new ensembles of LDPC codes, although any specific Kite code can be viewed as a known code. In
Section~\ref{sec3}, we analyze the maximum likelihood~(ML) decoding performance of the
Kite codes by applying a refined Divsalar bound to the ensemble
weight enumerating function~(WEF). In Section~\ref{sec4}, a greedy optimizing algorithm
to design Kite codes is introduced. We also discuss the optimization algorithm by using density evolutions. In Section~\ref{sec5}, we
present the serial concatenations of RS codes and Kite codes~(for
brevity, RS-Kite codes), followed by the performance evaluation of
RS-Kite codes and construction examples in Section~\ref{sec6}.
Section~\ref{conclusion} concludes this paper.

{\bf Notation:} A random variable is denoted by an upper case letter
$X$, whose realization is denoted by the corresponding lower case
letter $x$. We use $P_{X}(x)$ to represent the probability mass
function for a discrete variable $X$ or probability density function
for a continuous variable $X$. We use ${\rm Pr}\{A\}$ to represent
the probability that the event $A$ occurs.

\section{Kite Codes}\label{sec2}

\subsection{Prefix Rateless LDPC Codes}
Let $\mathbb{F}_q$ be the finite field with $q$ elements. Let
$\mathbb{F}_q^{\infty}$ be the set of all infinite sequences over
$\mathbb{F}_q$. As we know, $\mathbb{F}_q^{\infty}$ is a linear
space under the conventional sequence addition and scalar
multiplication. A {\em rateless linear code} $\mathcal{C}[\infty,k]$
is defined as a $k$-dimensional subspace of $\mathbb{F}_q^{\infty}$.
Let $\mathcal{C}[n,k]$ be the prefix code of length $n$ induced by
$\mathcal{C}[\infty,k]$, that is,
 \begin{equation}\label{prefix-code}
    \mathcal{C}[n,k] \stackrel{\Delta}{=}\left\{
    {\underline c}[n]\stackrel{\Delta}{=}(c_0, c_1, \cdots, c_{n-1})\mid
    {\underline c}[n]\textrm{ is a prefix of some infinite
    sequence } {\underline c} \in \mathcal{C}[\infty,k]\right\}.
\end{equation}
Clearly, $\mathcal{C}[n,k]$ is a linear code. If $\mathcal{C}[k,k]$
(hence $\mathcal{C}[n,k], n > k$) has dimension $k$, we call
$\mathcal{C}[\infty,k]$ a {\em prefix rateless linear code}.
Equivalently, a prefix linear code can be used as a systematic code
by treating the initial $k$ bits as information bits. Furthermore,
if all $\mathcal{C}[n,k]$ have low-density parity-check matrices, we
call $\mathcal{C}[\infty,k]$ a {\em prefix rateless low-density
parity-check~(PRLDPC) code}.

\subsection{Kite Codes}
\begin{figure}
\centering
\includegraphics[width=10cm]{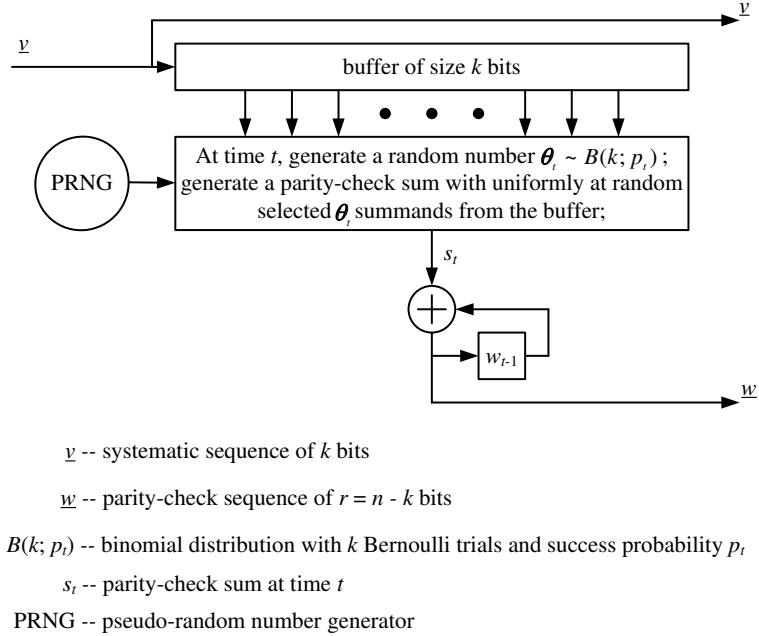}
\caption{A systematic encoding algorithm for the proposed Kite
code.} \label{inner-encoding}
\end{figure}

We now present a special class of PRLDPC codes, called {\em Kite
codes}. An ensemble of Kite codes, denoted by
$\mathcal{K}[\infty,k;\underline{p}]$, is specified by its dimension
$k$ and a real sequence $\underline p = (p_0, p_1, \cdots, p_{t},
\cdots)$ with $0< p_t < 1$ for $t\ge 0$. For
convenience, we call this sequence {\em p-sequence}.
The encoder of a Kite code consists of a buffer of size $k$, a
pseudo-random number generator~(PRNG) and an accumulator, as shown
in Fig.~\ref{inner-encoding}. We assume that the intended
receivers know exactly the {\it p}-sequence and the seed of the
PRNG.

Let $\underline v = (v_0, v_1, \cdots, v_{k-1})$ be the binary
sequence to be encoded. The corresponding codeword is written as 
$\underline c = (\underline v, \underline w)$, where 
$\underline w = (w_0, w_1, \cdots, w_t, \cdots)$ is the parity-check 
sequence. The task of the encoder is to compute $w_t$ for any $t \geq 0$. The encoding
algorithm is described as follows.

\begin{algorithm}{(Recursive encoding algorithm for Kite
codes)}\label{Kencoding}
\begin{enumerate}
    \item {\em Initialization:} Initially, set $w_{-1} = 0$.
     Equivalently, set the initial state of the accumulator to be zero.

    \item {\em Recursion:} For $t = 0, 1, \cdots, T-1$ with $T$ as large as required, compute $w_t$
    recursively according to the following procedures.

    {\em \underline {Step 2.1}}: At time $t \geq 0$, sampling from $k$ independent identically distributed~(i.i.d.) binary random
    variables $\underline H_t = (H_{t, 0}, H_{t, 1}, \cdots, H_{t, k-1})$, where
        \begin{equation}\label{binaryvariable}
            {\rm Pr}\{H_{t, i} = h_{t, i}\} = \left\{\begin{array}{cc}
              p_t, & h_{t,i} = 1 \\
              1-p_t, & h_{t,i} = 0 \\
            \end{array} \right.
        \end{equation}
    for $0\leq i < k$;

    {\em \underline {Step 2.2}}: Compute the $t$-th parity-check bit by $w_t = w_{t-1} + \sum_{0\leq i < k} h_{t, i}v_i~{\rm mod}~2$.
\end{enumerate}
\end{algorithm}

{\bf Remark.} Algorithm 1 can also be implemented by
first sampling from the binomial distribution $B(k; p_t)$ to
obtain an integer $\theta_t$ and then choosing $\theta_t$ summands
uniformly at random from the buffer to obtain a parity-check sum
$s_t$, which is utilized to drive the accumulator, as shown in
Fig.~\ref{inner-encoding}.

For convenience, the prefix code $\mathcal{K}[n,k]$ of a Kite code
$\mathcal{K}[\infty,k;\underline{p}]$ is also called a Kite code.
A Kite code $\mathcal{K}[n,k]$ for $n\geq k$ is a systematic
linear code with $r\stackrel{\Delta}{=}n-k$ parity-check bits. We
can write its parity-check matrix $H$ as
\begin{equation}\label{parity-check-matrix}
H = \left(H_v, H_w\right),
\end{equation}
where $H_v$ is a matrix of size $r\times k$ that corresponds to
the information bits, and $H_w$ is a square matrix of size
$r\times r$ that corresponds to the parity-check bits. By
construction, we can see that the sub-matrix $H_v$ is a
random-like binary matrix whose entries are governed by the {\it p}-sequence and
the initial state of the PRNG. In contrast, the square matrix
$H_w$ is a dual-diagonal matrix~(blanks represent zeros)

\begin{equation}\label{parity-check-matrix-right}
    H_w = \left(\begin{array}{ccccc}
      1         &           &          &           &        \\
      1         & 1         &          &           &        \\
                & 1         & \ddots   &           &        \\
                &           & \ddots   & 1         &        \\
                &           &          & 1         &1       \\
    \end{array}\right).
\end{equation}

\begin{figure}
\centering
\includegraphics[width=10cm]{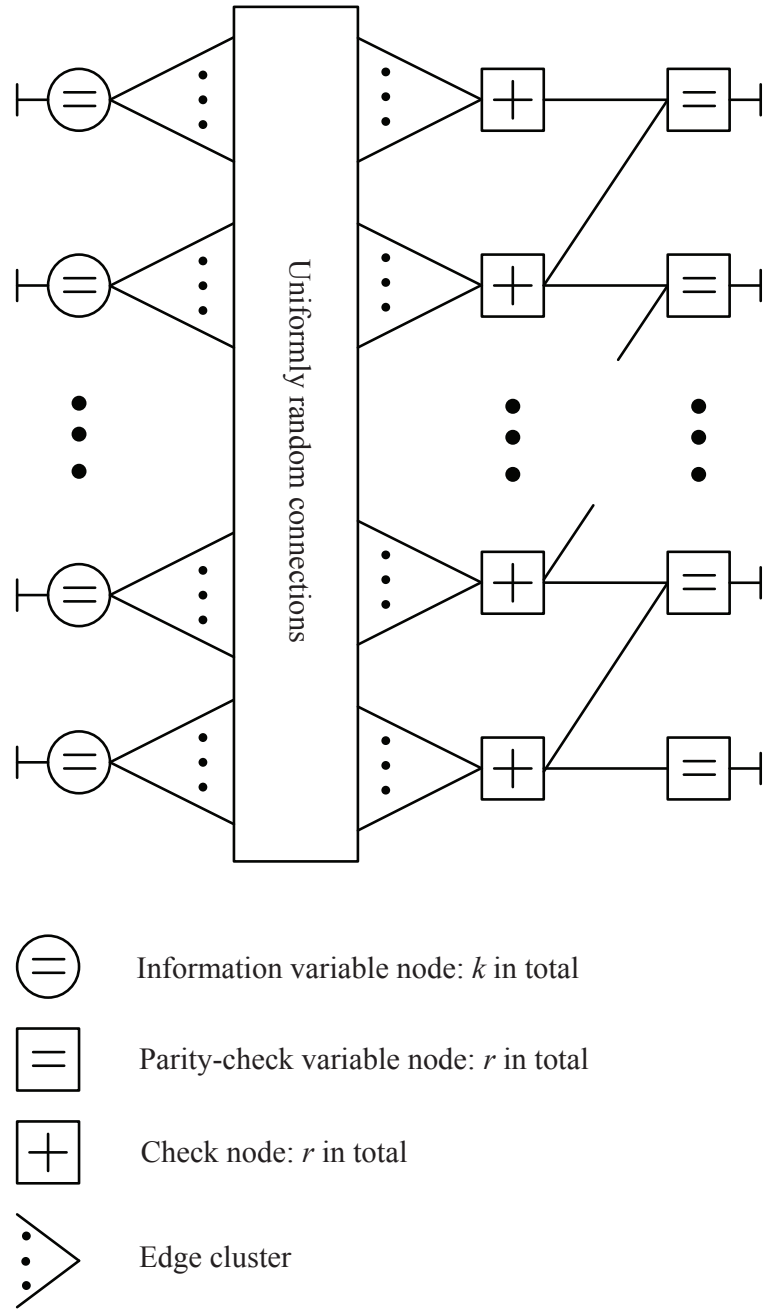}
\caption{A normal graph of a Kite code.} \label{inner-factor}
\end{figure}

Since the density of the parity-check matrix is dominated by the
$p$-sequence, we can construct Kite codes as LDPC codes by
choosing $p_t \ll 1/2$. If so, the receiver can
perform the iterative sum-product decoding
algorithm~\cite{Gallager62}. The iterative sum-product algorithm
can be described as a message passing/processing algorithm on
factor graphs~\cite{Kschischang01}. A Forney-style factor
graph~(also called a normal graph~\cite{Forney01}) representation
of $\mathcal{K}[n,k]$ is shown in Fig.~\ref{inner-factor}.

Now we assume that the coded bit $c_t$ at time $t\geq 0$ is
modulated in a BPSK format and
transmitted over the AWGN channel.
With this assumption, an intended receiver observes a noisy version
$y_t = x_t+z_t$, where $x_t = 1-2c_t$ and $z_t$ is an AWGN sample
at time $t$. The noise variance $\sigma^2$ is assumed to be known
at the receiver. We also assume that the intended receiver has a
large enough buffer to store the received sequence for the purpose
of decoding.

Once a noisy prefix~(of length $n$) ${\underline y}[n]$ is
available, the receiver can perform the iterative sum-product
decoding algorithm to get an estimate of the transmitted codeword
${\underline c}[n]$. The decoding is said to be {\em successful}
if the decoder can find an estimate $\hat{\underline c}[n]$ within
$J$~(a preset maximum iteration number) iterations satisfying all
the parity-check equations specified by $H$. If not such a case,
the receiver may collect more noisy bits to try the decoding
again.
If $n$ is the length of the prefix code that can be decoded
successfully, we call $k / n $ {\em decoding rate} rather than
{\em coding rate}. The decoding rate is a random variable, which
may vary from frame to frame. However, the expectation of the
decoding rate cannot exceed the corresponding channel
capacity~\cite{Shamai07}. Given a signal-to-noise-ratio~(SNR) of
$1/\sigma^2$, the gap between the average decoding rate and the
channel capacity can be utilized to evaluate the performance of
the proposed Kite code. A rateless code is called {\em universal}
if it is ``good" in a wide range of SNRs. It is
reasonable to guess that the universality of the proposed Kite
code for given $k$ can be improved by properly choosing the
$p$-sequence.

\subsection{Relations Between Kite Codes and Existing Codes}

The proposed Kite codes are different from existing iteratively decodeable codes.
\begin{enumerate}
  \item The Kite codes are different from general LDPC codes. An ensemble of Kite codes is {\em systematic}, {\em rateless} and characterized by the $p$-sequence, while an ensemble of general LDPC codes is usually non-systematic and characterized by two degree distributions $\lambda(x) = \sum_i \lambda_i x^{i-1}$ and $\rho(x) = \sum_i \rho_i x^{i-1}$, where
        $\lambda_i$~($\rho_i$) represents the fraction of edges emanating from variable (check) nodes of degree $i$~\cite{Luby97}\cite{Richardson01}.
  \item The Kite codes can be considered as serially concatenated codes with systematic low-density generator-matrix~(LDGM) codes~\cite{Garcia-Frias03}
        as outer codes and an accumulator as inner code. However, different from conventional serially concatenated codes, the inner code takes only the parity-check bits from the
        outer codes as input bits. An another difference is as follows. The generator matrices $H_v^{\bf T}$~(transpose of $H_v$) of outer LDGM codes may have infinite columns with random weights governed by the $p$-sequence rather than degree polynomials.
  \item The Kite codes are also similar to the generalized irregular repeat-accumulate~(GeIRA) codes~\cite{Jin00}\cite{Yang04}\cite{Liva05}\cite{Abbasfar07}. However, GeIRA codes are usually
        specified by the repetition multiplicities of each information bits and the uniform interleaver.
  \item As a rateless coding method, the proposed Kite codes are different from LT-codes and Raptor codes. For LT-codes and Raptor codes, coded bits are generated independently according to a
        time-independent degree distribution; while for Kite codes, parity-check bits are generated dependently. The Kite codes are more similar to the codes proposed in~\cite{Yuan10}, which are designed for erasure channels and specified by degree polynomials.
\end{enumerate}

As an ensemble of codes, the proposed Kite codes are new. The main feature of Kite codes is the use of the $p$-sequence instead of degree distributions to define the ensemble. This has brought at least two advantages. Firstly, as shown in Sec.~\ref{sec3}, the weight enumerating function~(WEF) of the ensemble can be easily calculated, implying that the ML decoding performance of Kite codes can be analyzed. Secondly, as shown in Sec.~\ref{sec4}, Kite codes can be designed by a greedy optimization algorithm which consists of several one-dimensional search rather than high-dimensional differential evolution algorithms~\cite{Richardson01a}~\cite{Price97}.

\section{Maximum Likelihood Decoding Analysis of Kite Codes}
\label{sec3}
\subsection{Weight Enumerating Function of the Ensemble of Kite Codes}
We can
define the {\em input-redundancy weight enumerating
function}~(IRWEF) of an ensemble of~(prefix) Kite codes with dimension $k$ and length $n$
as~\cite{Benedetto96}
\begin{equation}\label{IRWEFA}
    A(X, Z) \stackrel {\Delta}{=} \sum\limits_{i,j} A_{i,j}X^i Z^j,
\end{equation}
where $X, Z$ are two dummy variables and $A_{i,j}$ denotes the
ensemble average of the number of codewords $\underline c =
(\underline v, \underline w)$ consisting of an input information
sequence $\underline v$ of Hamming weight $i$ and a parity check
sequence $\underline w$ of Hamming weight $j$.

Let $\underline v^{(\ell)}$ be an input sequence consisting of
$\ell$ ones followed by $k-\ell$ zeros. That is, $\underline
v^{(\ell)} \stackrel{\Delta} {=} (\underbrace{1 \cdots 1}_\ell
\underbrace{0 \cdots 0}_{k-\ell})$. Let $\underline w^{(\ell)}$ be
the resulting parity-check sequence, which is a sample of a random
sequence $\underline W^{(\ell)}$ depending on the choice of the
parity-check matrix. Given $\underline v^{(\ell)}$, the
probability ${\rm Pr}\{W_t^{(\ell)}=1\}$ can be determined
recursively as follows.

Firstly, note that the random binary sum~(see Fig.~\ref{inner-encoding} for
the notation $s_t$) $S_t^{(\ell)} = \sum\limits_{0\leq i \leq
\ell-1} H_{t, i}$. So $p_t^{(\ell)} \stackrel{\Delta}{=}{\rm
Pr}\{S_t^{(\ell)} = 1\}$ can be calculated recursively as
$p_t^{(\ell)} = p_t^{(\ell-1)}(1-p_t) + (1-p_t^{(\ell-1)})p_t$ for
$\ell>0$ whereby $p_t^{(0)}$ is initialized as zero. Secondly, we
note that the sequence $\underline W^{(\ell)}$ is a Markov process
with the following time-dependent transition probabilities

\begin{equation}\label{probtrans}
\begin{array}{ll}
  {\rm Pr}\{W_t^{(\ell)} = 0\mid W_{t-1}^{(\ell)}=0\} = 1-p_t^{(\ell)}, & {\rm Pr}\{W_t^{(\ell)} = 1\mid W_{t-1}^{(\ell)}=0\} = p_t^{(\ell)}, \\
  {\rm Pr}\{W_t^{(\ell)} = 0\mid W_{t-1}^{(\ell)}=1\} = p_t^{(\ell)}, & {\rm Pr}\{W_t^{(\ell)} = 1\mid W_{t-1}^{(\ell)}=1\} = 1-p_t^{(\ell)}. \\
\end{array}
\end{equation}

\vspace{0.5cm}

We have the following two propositions.

\begin{figure}
\centering
\includegraphics[width=10cm]{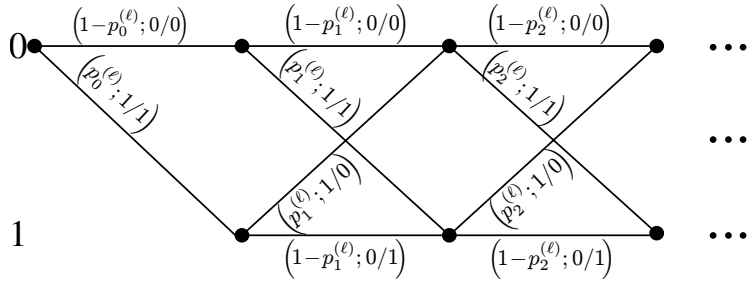}
\caption{The trellis representation of the random parity-check
sequence $\underline{W}^{(\ell)}$. To calculate the ensemble WEF,
we assign to each branch a metric ${\rm Pr}\{S_t^{(\ell)} =
s_t\}\cdot Z^{w_t}$, where $s_t/w_t$ are the input/output
associated with the branch, respectively.}\label{alpha-bcjr}
\end{figure}

\begin{proposition}\label{proposition1} Let $A^{(\ell)}(Z)$ be the
ensemble average weight enumerating function of $\underline
W^{(\ell)}$. Let $\alpha_t(Z; w)$ be the ensemble average weight
enumerating function of the prefix sequence $\underline
W^{(\ell,t)} = (W^{(\ell)}_0, W^{(\ell)}_1, \cdots, W^{(\ell)}_t)$
ending with $W^{(\ell)}_t = w$ at time $t$. Then $A^{(\ell)}(Z)$
can be calculated recursively by performing a forward
trellis-based algorithm (see Fig. \ref{alpha-bcjr} for the trellis
representation) over the polynomial ring.
\begin{itemize}
    \item Initially, set $\alpha_{-1}(Z; 0) = 1$ and $\alpha_{-1}(Z; 1) = 0$;

    \item For $t \geq  0$,
    \begin{eqnarray*}
      \alpha_t(Z; 0) &=&  (1-p_t^{(\ell)}) \alpha_{t-1}(Z; 0) + p_t^{(\ell)}\alpha_{t-1}(Z; 1)\\
      \alpha_t(Z; 1) &=&   p_t^{(\ell)} Z \cdot \alpha_{t-1}(Z; 0) + (1-p_t^{(\ell)})Z \cdot \alpha_{t-1}(Z; 1)
    \end{eqnarray*}

    \item At time $r-1$, we have $A^{(\ell)}(Z) = \alpha_{r-1}(Z; 0) + \alpha_{r-1}(Z; 1)$.
\end{itemize}
\end{proposition}

\begin{IEEEproof} The algorithm is similar to the trellis algorithm over polynomial rings for computing the weight enumerators of paths~\cite{McEliece96}.
The difference is that each path here has a probability, which can be incorporated into the recursions by assigning to each branch the corresponding
transition probability as a factor of the branch metric.
\end{IEEEproof}

\vspace{0.5cm}

\begin{proposition}\label{proposition2}
$A(X, Z) = \sum\limits_{0 \leq \ell \leq k}
\left(\begin{array}{c}
  k \\
  \ell \\
\end{array}\right)X^\ell A^{(\ell)}(Z)$.
\end{proposition}

\begin{IEEEproof} By the
construction, we can see that the columns of $H_v$ are identically
independent distributed. This implies that $A^{(\ell)}(Z)$ depends
only on the Hamming weight $\ell$ of $\underline{v}$ but not on the
locations of the $\ell$ ones in the sequence $\underline{v}$.
\end{IEEEproof}

\subsection{Refined Divsalar Bound}
We now proceed to discuss the ML decoding performance of Kite codes.
Consider the prefix code $\mathcal{K}[n,k;\underline{p}]$. Assume
that the all-zero codeword $\underline{c}^{0}[n]$ is transmitted
and $\underline{y}[n]$ is the received vector with
$y_t=1-2c_t+z_t$ and $z_t$ is an AWGN sample with zero mean and
variance $\sigma^2$. The probability of ML decoding error can be
expressed as
\begin{equation}\label{error-prob}
 {\rm Pr}\left\{E\right\} =
 {\rm Pr}\left\{\bigcup_{d}E_d\right\},
\end{equation}
where $E_d$ is the event that there exists at least one codeword of
weight $d$ that is nearer to $\underline{y}[n]$ than
$\underline{c}^{0}[n]$.

Let $S_d = \sum\limits_{i,j: i+j = d}A_{i, j}$ for all $0\leq
d\leq n$. Divsalar derived a simple upper
bound~\cite{Divsalar99}\cite{Divsalar03}
\begin{equation}\label{DivsalarBound}
    {\rm Pr}\{E_d\}\leq \min\left\{e^{-nE(\delta, \beta, \gamma)}, S_dQ\left(\sqrt{2d\gamma}\right)\right\},
\end{equation}
where
\begin{equation}\label{DivsalarExp}
    E(\delta, \beta, \gamma) = -r_n(\delta) + \frac{1}{2}\ln\left(\beta +
    (1-\beta)e^{2r_n(\delta)}\right) + \frac{\beta\gamma\delta}{1-(1-\beta)\delta}
\end{equation}
and $\gamma = \frac{E_s}{2\sigma^2}$, $\delta = d/n$, $r_n(\delta) =
\frac{\ln S_d}{n}$ and
\begin{equation}\label{Beta}
    \beta = \sqrt{\frac{\gamma(1-\delta)}{\delta}\frac{2}{1-e^{-2r_n(\delta)}} +
    \left(\frac{1-\delta}{\delta}\right)^2\left[(1+\gamma)^2-1\right]}
    - \frac{1-\delta}{\delta}(1+\gamma).
\end{equation}
From~(\ref{DivsalarBound}) and the union bound, we have
\begin{equation}\label{oribound}
 {\rm Pr}\left\{E\right\} \leq \sum_d
 \min\left\{e^{-nE(\delta, \beta, \gamma)}, S_dQ\left(\sqrt{2d\gamma}\right)\right\}.
\end{equation}

The question is, how many terms do we need to count in the above
bound? If too few terms are counted, we will obtain a lower
bound of the upper bound, which may be neither an upper bound nor a
lower bound; if too many are counted, we need pay more effort to
compute the distance distribution and we will obtain a loose upper
bound. To get a tight upper bound, we may determine the terms by
analyzing the facets of the Voronoi region of the codeword
$\underline{c}^{0}[n]$, which is a difficult task for a general
code. We have proposed a technique to reduce the number of terms~\cite{Ma10}.
For completeness, we include a brief description of the technique here. The basic idea is to limit the competitive candidate codewords by
using the following suboptimal algorithm.

\begin{algorithm}{(A list decoding algorithm for the purpose of
performance analysis)}\label{subDEC}
\begin{enumerate}
  \item[S1.] Make hard decisions, for $0\leq t \leq n-1$,
\begin{equation}
 \hat{y}_t = \left\{\begin{array}{cc}
                      0, & y_t > 0 \\
                      1, & y_t \leq 0
                    \end{array}
 \right..
\end{equation}
Then the channel $c_t\rightarrow \hat{y}_t$ becomes a memoryless
binary symmetric channel~(BSC) with cross probability
\begin{equation}\label{pBSC}
p_{BSC}=Q\left(\frac{1}{\sigma}\right)\stackrel{\Delta}{=}\int
_{1/\sigma}^{+\infty}\frac{1}{\sqrt{2\pi}}e^{-\frac{z^2}{2}}\,{\rm
d}z.
\end{equation}

  \item[S2.] List all codewords within the Hamming sphere with center
at $\underline{\hat{y}}$ of radius $d^*\geq 0$. The resulted list is
denoted as $\mathcal{L}_{\underline{y}}$.
  \item[S3.] If $\mathcal{L}_{\underline y}$ is empty, report a decoding error; otherwise, find the codeword $\underline{c}^*\in
\mathcal{L}_{\underline{y}}$ that is closest to $\underline{y}[n]$.
\end{enumerate}

\end{algorithm}

Now we define
\begin{equation}\label{Region}
    \mathcal{R} \stackrel{\Delta}{=} \left\{\underline y | {\underline c}^0[n] \in \mathcal{L}_{\underline y}\right\}.
\end{equation}
In words, the region $\mathcal{R}$ consists of all those $\underline y[n]$ having at most $d^*$ non-positive components.
The decoding error occurs in two cases under the assumption that the all-zero codeword ${\underline c}^0[n]$ is transmitted.

\begin{figure}
\centering
\includegraphics[width=10cm]{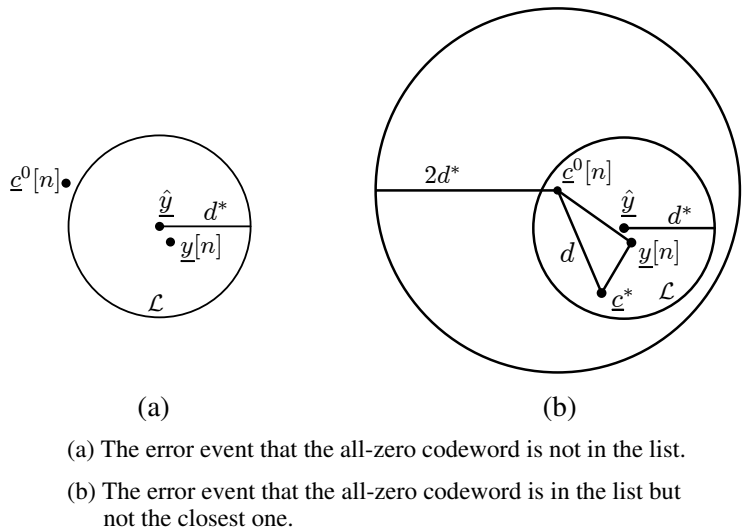}
\caption{Graphical illustrations of the decoding error events.}
\label{Error}
\end{figure}

{\em Case 1}. The all-zero codeword is not in the list $\mathcal{L}_{\underline y}$~(see Fig.~\ref{Error}~(a)), that is,  $\underline{y}[n]\notin \mathcal{R}$, which means that at least $d^* +1$ errors occur over the BSC. This probability is
\begin{equation}\label{Case1}
 {\rm Pr}\left\{\underline{y}[n]\notin \mathcal{R}\right\} = \sum_{t=d^*
 +1}^{n} \binom{n}{t} p_{BSC}^t(1-p_{BSC})^{n-t}.
\end{equation}

{\em Case 2}. The all-zero codeword is in the list
$\mathcal{L}_{\underline y}$, but is not the closest one~(see
Fig.~\ref{Error}~(b)), which is equivalent to the event $\left\{E,
\underline{y}[n] \in \mathcal{R} \right\}$. Since all codewords in the list $\mathcal{L}_{\underline{y}}$ are at
most $2d^*$ away from the all-zero codeword, this probability is
upper-bounded by
        \begin{eqnarray}\label{B1_Bound}
          {\rm Pr} \left\{E,\underline{y}[n]\in \mathcal{R} \right\}  &\leq& {\rm Pr} \left\{\bigcup_{d\leq 2d^*} E_d,\;\; \underline{y}[n]\in \mathcal{R} \right\}\label{GFBT1a}\\
           &\leq& {\rm Pr} \left\{\bigcup_{d\leq 2d^*} E_d,\;\;\underline{y}[n]\in \mathbb{R}^{n}\right\}\label{GFBT1b}\\
           &\leq&\sum_{d\leq 2d^*}{\rm Pr}\left\{E_d\right\}\label{B1}\\
           &\leq&\sum_{d\leq 2d^*}\min\left\{e^{-nE(\delta, \beta, \gamma)}, S_dQ\left(\sqrt{2d\gamma}\right)\right\}.\label{DivsalarBound1}
        \end{eqnarray}

Combining (\ref{Case1}) and (\ref{DivsalarBound1}) with Gallager's first bounding technique~(GFBT)~\cite{Sason06}
\begin{equation}\label{GFBT}
    {\rm Pr}\{E\} \leq {\rm Pr}\{E, {\underline y}[n] \in \mathcal{R}\} + {\rm Pr}\{{\underline y}[n] \notin \mathcal{R}\},
\end{equation}
we get an upper bound
\begin{equation}\label{newbound}
 {\rm Pr} \{E\} \leq \sum_{d \leq 2d^*} \min\left\{e^{-nE(\delta, \beta, \gamma)}, S_dQ\left(\sqrt{2d\gamma}\right)\right\} + \sum_{t> d^*} \binom{n}{t} p_{BSC}^t(1-p_{BSC})^{n-t}.
\end{equation}

To calculate the upper bound on bit-error probability, we need to
replace $S_d$ in (\ref{newbound}) by
\begin{equation}\label{WEF4B}
    S'_d = \sum\limits_{i, j: i+j = d}\frac{i}{k} A_{i,j}.
\end{equation}
Here we have used the fact that, in the case of ${\underline c}^0[n]\notin \mathcal{L}_{\underline y}$, the decoding error can contribute at most one to the bit-error
rate~(BER).

The modified upper bound in~(\ref{newbound}) has advantageous over
the original Divsalar bound~(\ref{oribound}). On one hand,
the modified upper bound in~(\ref{newbound}) is less complex than
the original Divsalar bound~(\ref{oribound}) for codes having no
closed-form WEFs since its computation only involves portion of the
weight spectrum up to $d\leq 2d^*$. On the other hand, if all $A_{i, j}$'s for $i+j\leq D$ are available,
we can get a tighter upper bound
\begin{equation}\label{optnewbound}
 {\rm Pr} \{E\} \leq \min_{d^* \leq D/2}\left\{\sum_{d \leq 2d^*} \min\left\{e^{-nE(\delta, \beta, \gamma)}, S_dQ\left(\sqrt{2d\gamma}\right)\right\} + \sum_{t> d^*} \binom{n}{t} p_{BSC}^t(1-p_{BSC})^{n-t}\right\},
\end{equation}
which can be tighter than the original Divsalar bound~(\ref{oribound})~(corresponding to $d^* = n$).

\begin{figure}
\centering
\includegraphics[width=10cm]{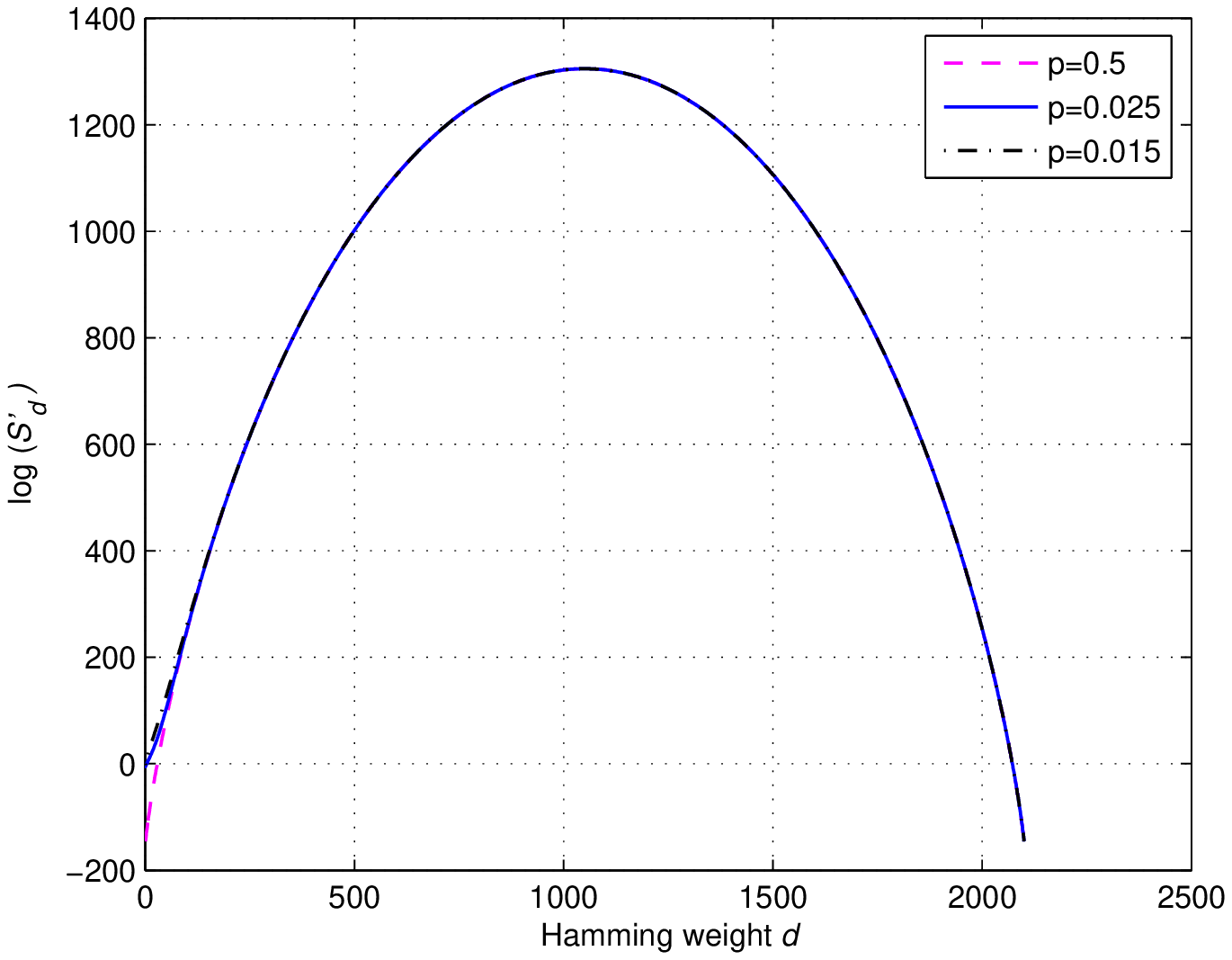}
\caption{Ensemble weight enumerating functions of
$\mathcal{K}[2100, 1890]$.} \label{WEF2100w}
\end{figure}

\begin{figure}
\centering
\includegraphics[width=10cm]{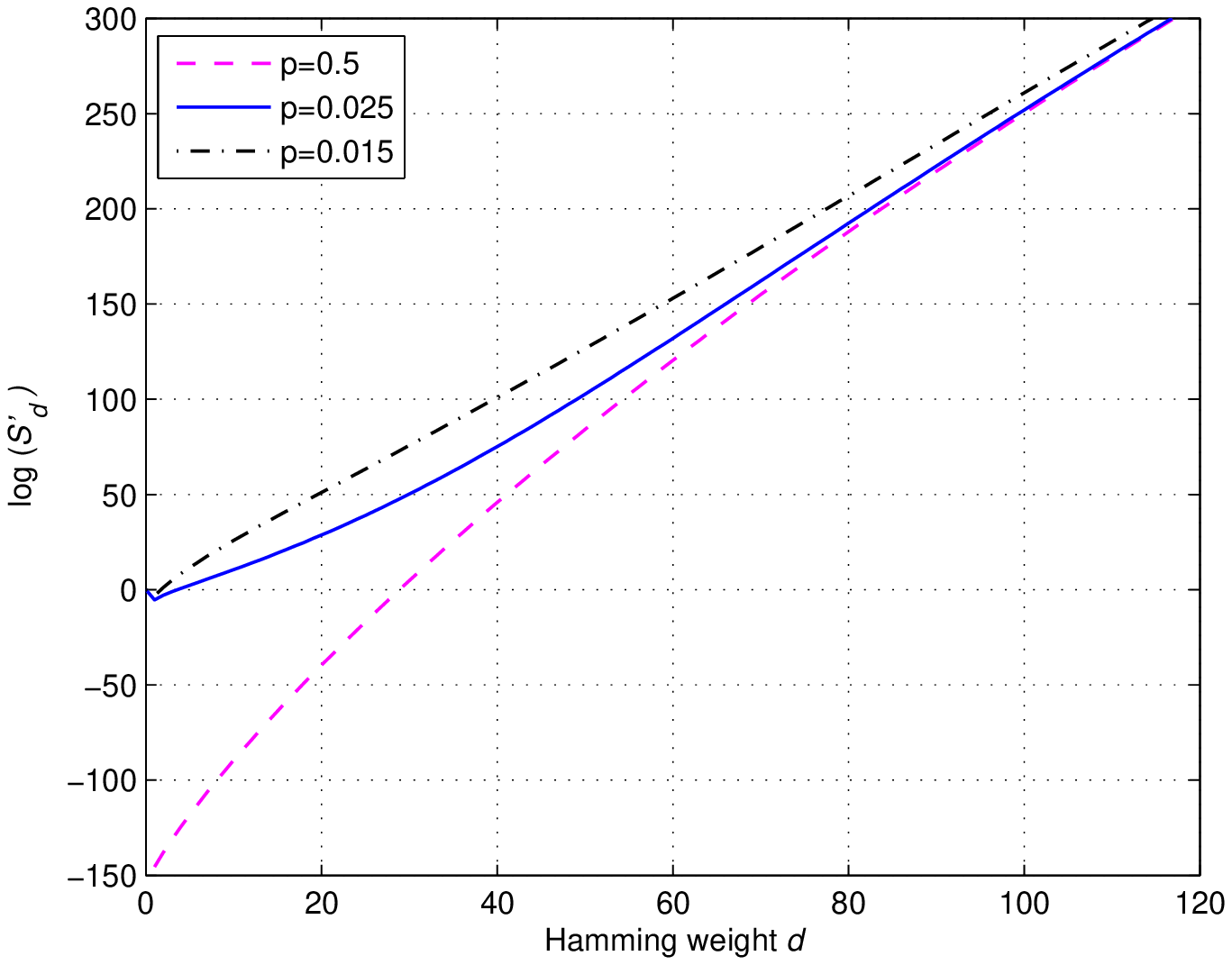}
\caption{Detailed ensemble weight enumerating functions of
$\mathcal{K}[2100, 1890]$.} \label{WEF2100p}
\end{figure}

\begin{figure}
\centering
\includegraphics[width=10cm]{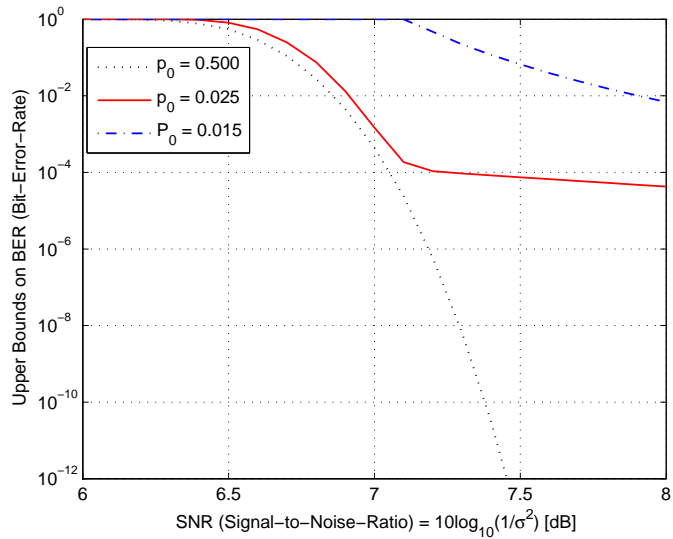}
\caption{Performance bounds on the ensemble Kite codes
$\mathcal{K}[2100, 1890]$.} \label{DBB2100}
\end{figure}

\begin{figure}
\centering
\includegraphics[width=10.0cm]{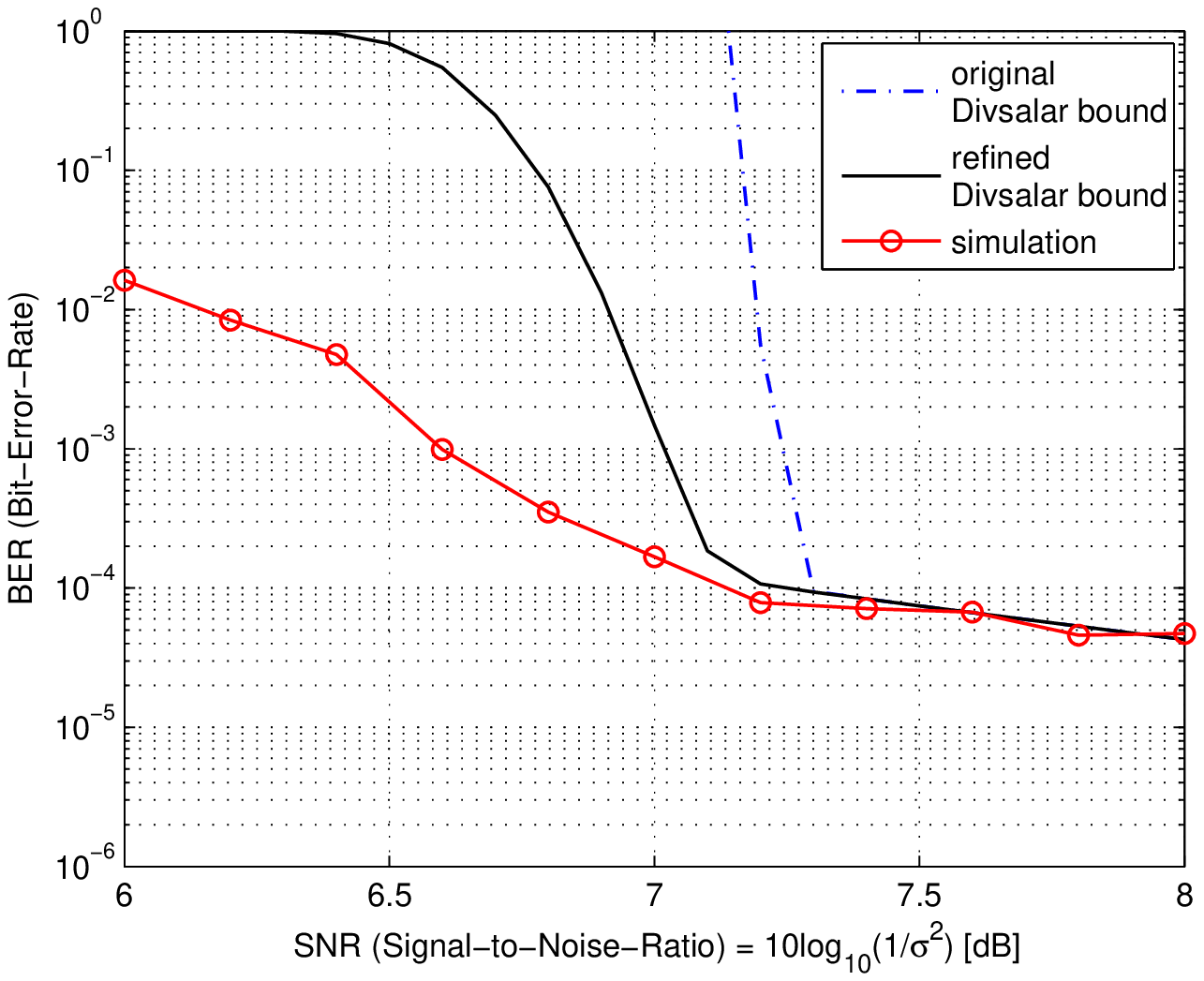}
\caption{Comparisons between bounds and simulation results for $p=0.025$.} \label{DBBSim2100}
\end{figure}

\subsection{Numerical Results}
We consider a Kite code $\mathcal{K}[2100, 1890]$. For simplicity,
set $p_t = p_0$ for all $t < 210$. We consider $p_0 = 0.5, 0.025$
and $0.015$, respectively. The WEFs $\{S'_d\}$ are shown in
Fig.~\ref{WEF2100w}. As we can see, the three curves match well in
the moderate-to-high-weight region. While, in the low-weight
region, they differs  much from each other, as shown in Fig.~\ref{WEF2100p}. As the
parameter $p_0$ increases to $0.5$, the resulting ensemble becomes
a uniformly random ensemble. The refined Divsalar bounds on BERs are shown in
Fig.~\ref{DBB2100}. Shown in Fig.~\ref{DBBSim2100} are the comparisons between bounds and simulation results for $p_0 = 0.025$. We have the following observations.
\begin{enumerate}
  \item The modified bound improves the original Divsalar bound especially in the low-SNR region.
  \item Under the assumption of the ML decoding, the performance degrades as the parameter $p_0$ decreases. There exists an error floor at BER around $10^{-4}$ for
        $p_0 = 0.025$. At the corner of the floor, the performance gap~(in terms of upper bounds) between the code with $p_0 = 0.025$ and the totally random code is less than $0.1$ dB.
  \item The iterative sum-product decoding algorithm delivers a curve that matches well with the performance bounds of the ML decoding in the high-SNR region.
\end{enumerate}

\section{Design of Kite Codes}
\label{sec4}
\subsection{Partition the $p$-Sequence According to Decoding Rates}
As we have observed in Section~\ref{sec3} that the choices of the
$p$-sequence have effect on the performance of Kite codes. On one
hand, to ensure with high probability that the resulting Kite
codes are iteratively decodable LDPC codes, we must choose the
$p$-sequence such that $p_t \ll 1/2$. On the other hand, to
guarantee the performance, the components of the $p$-sequence can
not be too small.

The task to optimize a Kite code is to select the whole
$p$-sequence such that all the prefix codes are good enough. This
is a multi-objective optimization problem and could be very
complex. For simplicity, we only consider the codes with rates not less than 0.1 and
simply group the $p$-sequence according to decoding
rates as follows
\begin{equation}\label{pSequence}
    p_t = \left\{\begin{array}{rl}
      q_9, & 0.9 \leq k/(t+k)<  1.0  \\
      q_8, & 0.8 \leq k/(t+k) <  0.9 \\
      q_7, & 0.7 \leq k/(t+k) <  0.8 \\
      q_6, & 0.6 \leq k/(t+k) <  0.7 \\
      q_5, & 0.5 \leq k/(t+k) <  0.6 \\
      q_4, & 0.4 \leq k/(t+k) <  0.5 \\
      q_3, & 0.3 \leq k/(t+k) <  0.4 \\
      q_2, & 0.2 \leq k/(t+k) <  0.3 \\
      q_1, & 0.1 \leq k/(t+k) <  0.2 \\
    \end{array}\right..
\end{equation}
Then the task to design a Kite code is to select the parameters
$\underline q = (q_9, q_8, \cdots, q_1)$.

\begin{figure}
\centering
\includegraphics[width=10cm]{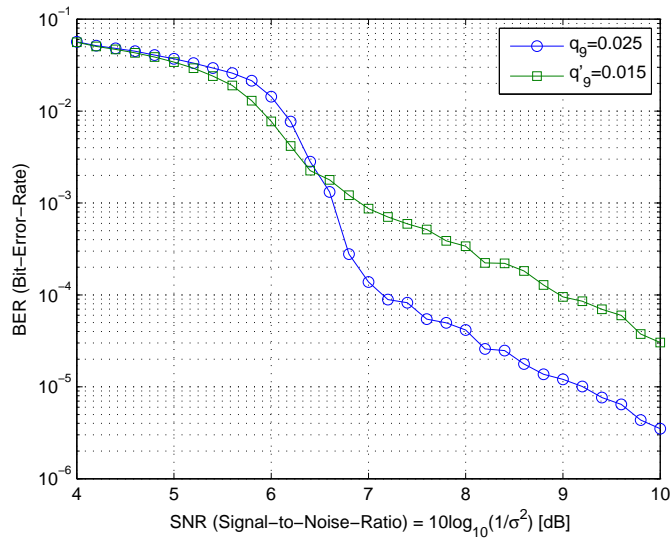}
\caption{BER performance of $\mathcal{K}[2100, 1890]$.}
\label{BER2100}
\end{figure}

\subsection{Greedy Optimizing Algorithms}

We use a greedy algorithm to optimize the parameters $\underline q$.
Firstly, we choose $q_9$ such that the prefix code
$\mathcal{K}[\lfloor k/0.9 \rfloor, k]$ is as good as possible.
Secondly, we choose $q_8$ with fixed $q_9$ such that the prefix code
$\mathcal{K}[\lfloor k/0.8 \rfloor, k]$ is as good as possible.
Thirdly, we choose $q_7$ with fixed $(q_9, q_8)$ such that the
prefix code $\mathcal{K}[\lfloor k/0.7 \rfloor, k]$ is as good as
possible. This process continues until $q_1$ is selected.

Let $q_{\ell}$ be the parameter to be optimized. Since the parameters
$q_j$ with $j > \ell$ have been fixed and the parameters $q_j$ with
$j < \ell$ are irrelevant to the current prefix code, the problem to
design the current prefix code then becomes a one-dimensional
optimization problem, which can be solved, for example, by the
golden search method~\cite{Rardin98}. What we need to do is to
make a choice between any two candidate parameters $q_\ell$ and
$q'_\ell$, which can be done~(with certain BER performance criterion) by at least two methods.
We take $k = 1890$ as an example to illustrate the main
idea.

\subsubsection{Simulation-Based Methods}
We can first simulate the BER versus
SNR curves and then make a choice between $q_\ell$ and $q'_\ell$. For example,
Fig.~\ref{BER2100} illustrates our simulations for
$\mathcal{K}[2100, 1890]$. If the target is $\textrm{BER} < 10^{-3}$,
we say that $q_9 = 0.025$ is superior to $q'_9 = 0.015$. At first
glance, such a simulation-based method could be very
time-consuming. However, it is fast due to the following two
reasons. Firstly, in this paper, our optimization target is set to
be $\textrm{BER} = 10^{-4}$, which can be reliably estimated without
large amounts of simulations. Secondly, our goal is not to
reliably estimate the performances corresponding to the two
parameters $q_\ell$ and $q'_\ell$ but to make a choice between these two
parameters. Making a choice between two design parameters, as an
example of {\em ordinal optimization} problem~\cite{Ho99}, can be
easily done by simulations.

\subsubsection{Density Evolution}
We can also make a choice between $q_\ell$ and $q'_\ell$ by using density evolution. As shown in Fig.~\ref{inner-factor}, the Kite code can be represented by a normal graph which contains three types of nodes: {\em information variable nodes}, {\em parity-check variable nodes} and {\em check nodes}, which are simply referred to as $A$-type nodes, $B$-type nodes and $C$-type nodes, respectively. Since the connections between $B$-type nodes and $C$-type nodes are deterministic, we only need to derive the degree distributions for the edges between $A$-nodes and $C$-nodes.

\textbf{Node degree distributions:} Assume that the parameters $q_j$ for $j > \ell$ have been optimized and fixed. The resulting parity-check matrix is written as $H^{(\ell+1)} = (H_v^{(\ell+1)}, H_w^{(\ell+1)})$, which is a random matrix of size $r^{(\ell+1)} \times (k + r^{(\ell+1)})$. In our settings, for example,
$r^{(9)} = \lfloor k / 0.9 \rfloor - k$, $r^{(8)} = \lfloor k / 0.8 \rfloor - k$, and so on.
Noting that $H^{(\ell)}$ is constructed from $H^{(\ell+1)}$ by adding $\delta = r^{(\ell)} - r^{(\ell+1)}$ rows, we may write it as
\begin{equation}\label{newH}
    H^{(\ell)} = (H_v^{(\ell)}, H_w^{(\ell)}) = \left(
                        \begin{array}{cc}
                          H_v^{(\ell +1)} & H_w^{(\ell +1)}  \\
                          H_v^{(\delta)} & H_w^{(\delta)} \\
                        \end{array}
                      \right).
\end{equation}
Since $H_w^{(\ell)}$ is deterministic, we only need to determine the degree distributions of nodes corresponding to $H_v^{(\ell)}$. For doing so, assume that the degree distributions of nodes corresponding to $H_v^{(\ell+1)}$ are, for $A$-type nodes,
\begin{equation}\label{Ax}
    \Lambda^{(\ell+1)}(x) = \sum_{0 \leq i \leq r^{(\ell+1)}} \Lambda_i^{(\ell+1)} x^i,
\end{equation}
and for $C$-type nodes,
\begin{equation}\label{Rx}
    R^{(\ell+1)}(x) = \sum_{0 \leq i \leq k} R_i^{(\ell+1)} x^i,
\end{equation}
where $\Lambda_i^{(\ell+1)}$~(resp. $R_i^{(\ell+1)}$) represents the fraction of $A$-type~(resp. $C$-type) nodes of degree $i$. In our settings, for example, $\Lambda_i^{(9)} = \binom{r^{(9)}}{i}q_9^i(1-q_9)^{r^{(9)}-i}$  for $0\leq i \leq r^{(9)}$ and $R_i^{(9)} = \binom{k}{i}q_9^i(1-q_9)^{k-i}$ for $0\leq i \leq k$. Since $\Lambda_0^{(9)} > 0$, there may exist nodes with no edges. However, such a probability can be very small for large $r^{(9)}$. It should also be pointed out that $R^{(\ell+1)}(x)$ only counts the edges of the $C$-type nodes connecting to the $A$-type nodes. For this reason, we call $R^{(\ell)}(x)$ the {\em left degree} distribution of $C$-nodes.

Let $\Lambda(x)=\sum_{0\leq i \leq \delta} \Lambda_i x^i$ and $R(x)=\sum_{0\leq i \leq k}R_i x^{i}$ be the degree distributions of nodes corresponding to $H_v^{(\delta)}$. Then we have
\begin{equation}
      \Lambda_i = \binom{\delta}{i}q_{\ell}^i(1-q_\ell)^{\delta-i}, \;\; 0\leq i \leq \delta
\end{equation}
and
\begin{equation}
      R_i = \binom{k}{i}q_\ell^i (1-q_\ell)^{k-i}, \;\; 0\leq i \leq k.
\end{equation}
Then it is not difficult to verify that
\begin{equation}
      \Lambda_i^{(\ell)} = \sum_{i = j + k} \Lambda_j \Lambda_k^{(\ell + 1)}, \;\; 0 \leq i \leq r^{(\ell)}
\end{equation}
and
\begin{equation}
      R_i^{(\ell)} = \frac{r^{(\ell+1)}R^{(\ell)}_i + \delta R_i}{r^{(\ell)}}, \;\; 0 \leq i \leq k.
\end{equation}

\textbf{Edge degree distributions:} Given $\Lambda^{(\ell)}(x)$ and $R^{(\ell)}(x)$~(which can be computed recursively), we can find the degree distributions
of edges between $A$-type nodes and $C$-type nodes as follows. Let $\lambda(x) = \sum_i \lambda_i x^{i-1}$ and $\rho(x) = \sum_i \rho_i x^{i-1}$ be the degree distributions of edges corresponding to $H_v^{(\ell)}$. We have
\begin{equation}
      \lambda_i = \frac{i\Lambda_i}{\sum_j j\Lambda_j}, \;\; 1\leq i \leq r^{(\ell)}
\end{equation}
and
\begin{equation}
      \rho_i = \frac{iR_i}{\sum_j jR_j}, \;\; 1\leq i \leq k.
\end{equation}

\textbf{Gaussian Approximation:}
The density evolution~\cite{Richardson01} is an algorithm that predicts the performance thresholds of random LDPC codes by tracing the probability density function~(pdf)
of the messages exchanging between different types of nodes under certain assumptions\footnote{Strictly speaking, the density evolution can not
be applied here to optimize the parameters because we are constructing codes with a fixed dimension $k$, the parameter $q_\ell$ is evidently dependent of the code dimension,
and the code graph is only semi-random.}. In the density evolution, we assume that the all-zero codeword is transmitted. It is convenient to assume that the input messages to the decoder
are initialized by the following log-likelihood ratios~(LLRs)
\begin{equation}
    M_0 = \log\frac{f_Y(y|+1)}{f_Y(y|-1)},
\end{equation}
where $f_Y$ is the conditional pdf of $Y$ given that $+1$ is transmitted. It has been proven~\cite{Richardson01} that, for the BPSK input and continuous output AWGN channel, $M_0$ is a Gaussian random variable having mean $\mu_0 = 2/\sigma^2$ and satisfying the symmetry condition $\sigma_0^2 = 2 \mu_0$. It was further shown~\cite{Chung01a} that all the intermediate messages produced during the iterative sum-product algorithm can be approximated by Gaussian variables or mixture of Gaussian variables satisfying the symmetry condition. This means that we need to trace only the means of the messages.

For Kite codes shown in Fig.~\ref{inner-factor}, the density evolution using Gaussian approximations are slightly different from that general LDPC codes.
To describe the algorithm more clearly, we introduce the following notation.

$~~~\mu_0$~~~the mean of initial messages from the channel;\\
$\mu_i^{(A\rightarrow C)}$~~~the mean of messages from $A$-type nodes of degree $i$ to $C$-type nodes;\\
$\mu^{(B\rightarrow C)}$~~~the mean of messages from $B$-type nodes to $C$-type nodes;\\
$\mu_i^{(C\rightarrow A)}$~~~the mean of messages from $C$-type nodes of {\em left degree} $i$ to $A$-type nodes;\\
$\overline{\mu}^{(C\rightarrow A)}$~~~the average mean of messages from $A$-type nodes to $C$-type nodes;\\
$\mu_i^{(C\rightarrow B)}$~~~the mean of messages from $C$-type nodes of {\em left degree} $i$ to $B$-type nodes;\\
$\overline{\mu}^{(C\rightarrow B)}$~~~the average mean of messages from $C$-type nodes to $B$-type nodes.

Slightly different from~\cite{Chung01a} but following~\cite{Moon05}, we define
\begin{equation}
    \phi(x) \stackrel{\Delta}{=} \left\{\begin{array}{ll}
                                   \frac{1}{\sqrt{4 \pi x}} \int^{+\infty}_{-\infty} \tanh(y/2) e^{-(y-x)^2/(4x)}dy, & x > 0 \\
                                   0, & x = 0
                                 \end{array}
    \right..
\end{equation}

We may use the following algorithm to predict the performance threshold for the considered parameter $q_\ell$ under the iterative sum-product decoding algorithm.
\begin{algorithm}(Density evolution using Gaussian approximations for Kite codes)
\begin{enumerate}
  \item \textit{Input:} The degree distributions of nodes $\Lambda(x)$, $R(x)$; the degree distributions of edges $\lambda(x)$, $\rho(x)$; target $T_b$ of BER and absolute difference $\Delta_b$ of BERs for two successive iterations;
  \item \textit{Initializations:} Set $\varepsilon = \int^{0}_{-\infty} \frac{1}{\sqrt{4 \pi \mu_0}} e^{-(y - \mu_0)^2/(4\mu_0)}dy$, $\overline{\mu}^{(C \rightarrow A)}=0$ and $\overline{\mu}^{(C \rightarrow B)}=0$;
  \item \textit{iterations} - Repeat:

        \emph{\underline{Step 3.1}:} From $A$-type nodes to $C$-type nodes,
                                     \begin{equation}
                                           \mu^{(A \rightarrow C)}_i  = \mu_0 + (i-1)\overline{\mu}^{(C \rightarrow A)}, 1\leq i \leq r;
                                     \end{equation}

        \emph{\underline{Step 3.2}:} From $B$-type nodes to $C$-type nodes,
                                     \begin{equation}
                                           \mu^{(B \rightarrow C)} = \mu_0 + \overline{\mu}^{(C \rightarrow B)};
                                     \end{equation}

        \emph{\underline{Step 3.3}:}
                                      \begin{enumerate}
                                           \item From $C$-type nodes to $A$-type nodes,
                                                 \begin{equation}
                                                        \mu^{(C \rightarrow A)}_j = \phi^{-1}\left(\phi^2(\mu^{(B \rightarrow C)})\left(\sum_{i=1}^{r}\lambda_i\phi(\mu_i^{(A \rightarrow C)})\right)^{j-1}\right), 1\leq j \leq k;
                                                 \end{equation}

                                                 \begin{equation}
                                                        \overline{\mu}^{(C \rightarrow A)}  = \sum\limits_{j=1}^{k} \rho_j \mu^{(C\rightarrow A)}_j;
                                                 \end{equation}

                                           \item From $C$-type nodes to $B$-type nodes,
                                                 \begin{equation}
                                                        \mu^{(C \rightarrow B)}_j = \phi^{-1}\left(\phi(\mu^{(B \rightarrow C)})\left(\sum_{i=1}^{r}\lambda_i\phi(\mu_i^{(A \rightarrow C)})\right)^{j}\right), 0\leq j \leq k;
                                                 \end{equation}

                                                 \begin{equation}
                                                       \overline{\mu}^{(C \rightarrow B)}  = \sum\limits_{j = 0}^{k} R_j \mu^{(C \rightarrow B)}_j;
                                                 \end{equation}
                                   \end{enumerate}
          \emph{\underline{Step 3.4}:} Make decisions, for $0\leq i \leq r$, define $\mu_i \stackrel{\Delta}{=} \mu_0 + i \overline{\mu}^{(C\rightarrow A)}$; compute
                                       \begin{eqnarray}
                                             \varepsilon_i = \int^{0}_{-\infty} \frac{1}{\sqrt{4\pi\mu_i}} e^{-(y-\mu_i)^2/(4\mu_i)} dy
                                       \end{eqnarray}
                                       and
                                       \begin{equation}
                                        \varepsilon' = \sum\limits_{i = 0}^{r} \Lambda_i\varepsilon_i;
                                       \end{equation}
                                       if $\varepsilon'\leq T_b$ or $|\varepsilon -\varepsilon'| \leq \Delta_b$, exit the iteration; else set $\varepsilon = \varepsilon'$ and go to \emph{\underline{Step 3.1}}.

\end{enumerate}
\end{algorithm}

{\bf Remark.} Note that we have ignored the effect caused by the margin of the subgraph consisting of $B$-type nodes and $C$-type nodes. Also note that, different from the density evolution for general LDPC codes, degree distributions of nodes are also involved here.

\hspace {3.5cm}

\begin{figure}
\centering
\includegraphics[width=10cm]{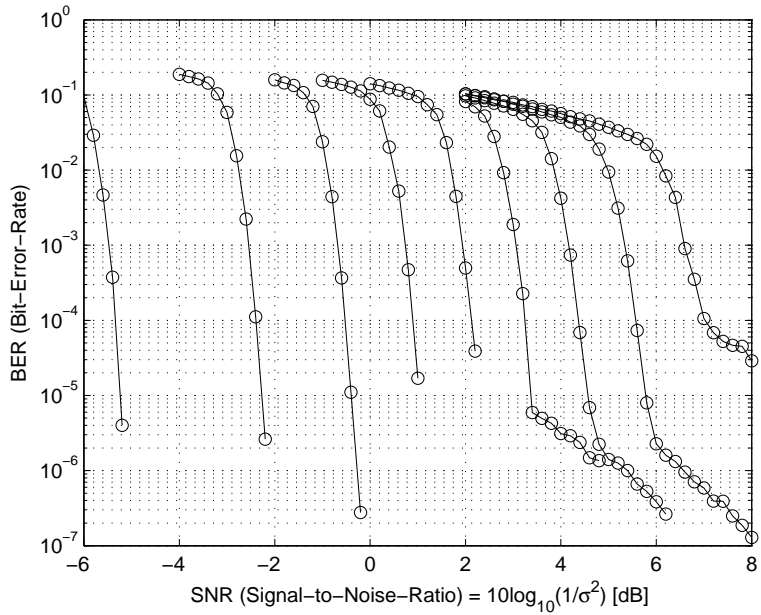}
\caption{Performance of the constructed Kite code with $k =1890$.
From left to right, the curves correspond to rates 0.1, 0.2, 0.3,
0.4, 0.5, 0.6, 0.7, 0.8, and 0.9, respectively.}
\label{BERKite1890}
\end{figure}

\subsection{Construction Examples}
In this subsection, we present a construction example. We take the
data length $k = 1890$. A Kite code $\mathcal{K}[\infty, 1890;
\underline p]$ can be constructed using the following $p$-sequence
\begin{equation}\label{pSequence1890}
    p_t = \left\{\begin{array}{rl}
      q_9 = 0.0249, & 0.9 \leq k/(t+k)<  1.0  \\
      q_8 = 0.0072, & 0.8 \leq k/(t+k) <  0.9 \\
      q_7 = 0.0045, & 0.7 \leq k/(t+k) <  0.8 \\
      q_6 = 0.0034, & 0.6 \leq k/(t+k) <  0.7 \\
      q_5 = 0.0021, & 0.5 \leq k/(t+k) <  0.6 \\
      q_4 = 0.0016, & 0.4 \leq k/(t+k) <  0.5 \\
      q_3 = 0.0010, & 0.3 \leq k/(t+k) <  0.4 \\
      q_2 = 0.0006, & 0.2 \leq k/(t+k) <  0.3 \\
      q_1 = 0.0004, & 0.1 \leq k/(t+k) <  0.2 \\
    \end{array}\right..
\end{equation}

The simulation results are shown in Fig.~\ref{BERKite1890}. For comparison, the SNR for the simulation at $\textrm{BER} = 10^{-4}$ and the SNR thresholds estimated by the Gaussian-approximation-based density evolution~(GA DE) are listed in the Table~\ref{SNRthreshold}. As we can see, the difference between the density evolution and  the simulation is less than 1 dB.

\begin{table}
  \centering
  \caption{The SNR for the density evolution and simulation at $\textrm{BER} = 10^{-4}$.}\label{SNRthreshold}
  \begin{tabular}{|r|r|r|}

  \cline{1-3} Code rate &  GA DE~(dB) &  Simulation~(dB)\\
  \cline{1-3}   0.9     &       6.49         &     7.0    \\
  \cline{1-3}   0.8     &       4.84         &    5.6     \\
  \cline{1-3}   0.7     &       3.65         &    4.4     \\
  \cline{1-3}   0.6     &       2.57         &    3.2     \\
  \cline{1-3}   0.5     &       1.48         &    2.1     \\
  \cline{1-3}   0.4     &       0.31         &    0.9     \\
  \cline{1-3}   0.3     &       -1.04        &    -0.5    \\
  \cline{1-3}   0.2     &       -2.84        &    -2.4    \\
  \cline{1-3}   0.1     &       -5.67        &    -5.3    \\
  \cline{1-3}

\end{tabular}
\end{table}

\section{Serial Concatenation of Reed-Solomon Codes and Kite Codes}
\label{sec5}
\subsection{Encoding}

\begin{figure}
\centering
\includegraphics[width=10cm]{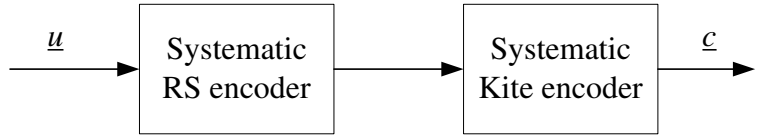}
\caption{Serial concatenation of RS codes and Kite Codes.}
\label{RSPRLDPCEncoding}
\end{figure}

Naturally, the proposed encoding method can be utilized to
construct fixed-rate codes. However, we found by simulation that,
in the range of moderate-to-high rates, the constructed fixed-rate
codes in this way suffer from error-floors at BER
around $10^{-4}$, as shown in Fig.~\ref{BERKite1890}. The error floor is caused by
the possibly existing the all-zero columns in the randomly generated parity-check matrices.
To lower-down the error-floor, we may insert some fixed-patterns into the matrices, as
proposed in~\cite{Bai11}. We also note
that, in the rateless coding scenario, the receiver must find a
way to ensure the correctness of the successfully decoded
codeword. This is trivial for erasure channels but becomes complex
for noisy channels. For erasure channels, a decoded codeword is
correct if and only if all information bits are recovered
uniquely. For AWGN channels, no simple way to ensure with
probability one the correctness of the decoded codeword at the
receiver. In order to lower down the error-floors and ensure with
high probability the correctness of the decoded codewords, we
employ the serially concatenated coding system proposed by
Forney~\cite{Forney66}. As shown in Fig.~\ref{RSPRLDPCEncoding},
the outer code is a systematic RS code of dimension
$K$ and length $N$ over $\mathbb{F}_{2^m}$, while the inner code
is a Kite code. Note that no interleaver is required between the
outer encoder and the inner encoder. For convenience, such a
concatenated code is called {\em an RS-Kite code}.

Let $\underline u$ be the binary data sequence of  length $\ell m
K$~($\ell > 0$). The encoding procedure is described as follows.
First, the binary sequence $\underline u$ is interpreted as a
$2^m$-ary sequence of length $\ell K$ and encoded by the outer
encoder, producing $\ell$ RS codewords. Then these $\ell$ RS
codewords are interpreted as a binary sequence $\underline v$ of
length $\ell m N$ and encoded using a Kite code, resulting in a
potentially infinite coded sequence $\underline c$.

\begin{figure}
\centering
\includegraphics[width=10cm]{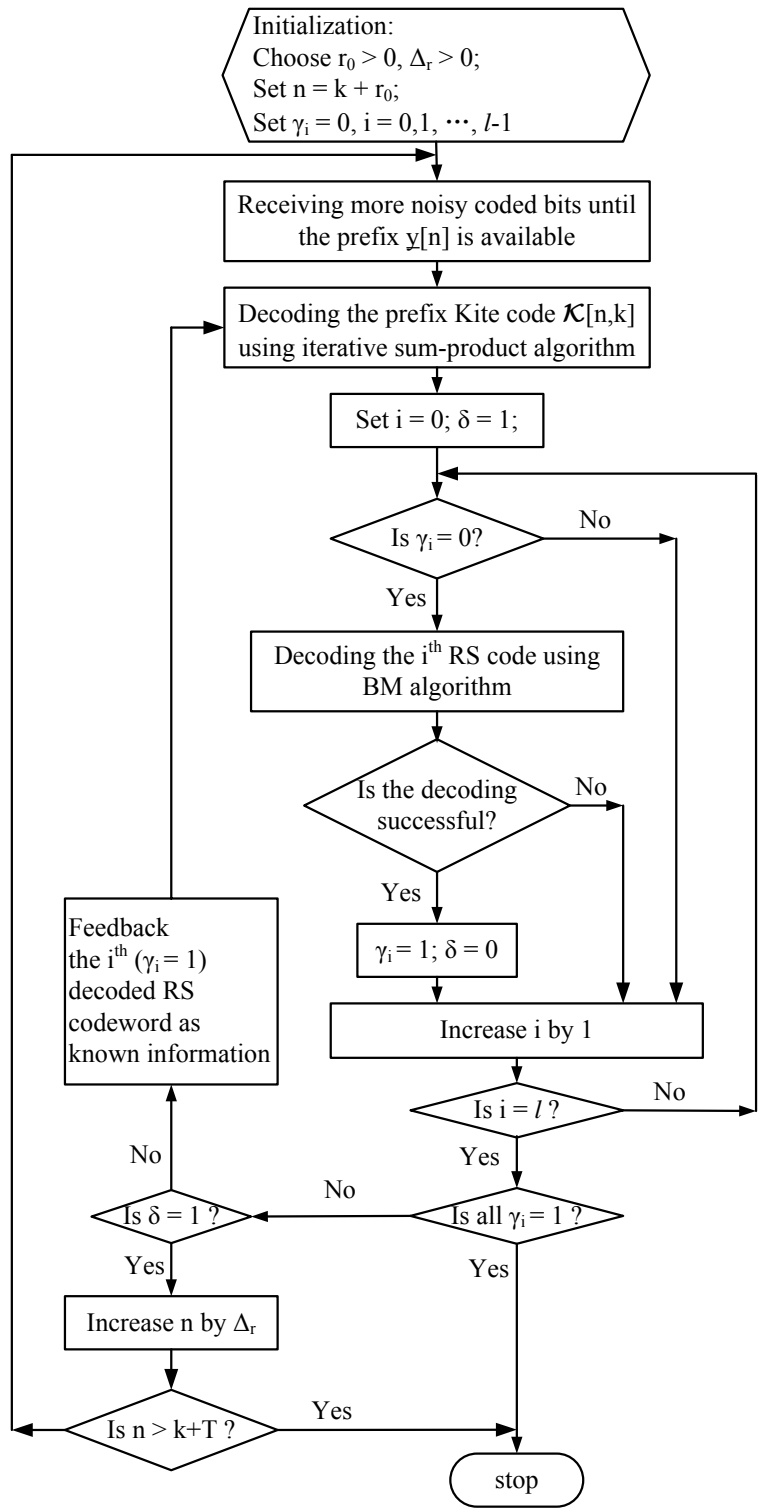}
\caption{A flowchart of the decoding algorithm for the proposed
serially concatenated codes.} \label{RSPRLDPCDecoding}
\end{figure}

\subsection{Decoding}

We will use Berlekamp-Massey~(BM) algorithm~\cite{Lin04} for the
outer decoder and iterative sum-product algorithm for the inner
decoder. As we know, if the number of errors in a noisy RS
codeword is not greater than $t_{max} \stackrel{\Delta}{=}
\lfloor(N-K)/2\rfloor$, the BM algorithm will find the correct
codeword. On the other hand, if the number of errors in a noisy RS
codeword is greater than $t_{max}$, the BM algorithm may claim a
decoding failure or make a miscorrection. We call the BM decoding
{\em successful} whenever the BM algorithm finds a codeword within
the Hamming sphere of radius $t_{max}$ around the received
``codeword" from the inner decoder. Since the miscorrection
probability can be made very small by properly setting the
parameters of the RS code~\cite{McEliece86}\cite{Sofair00}, we
will treat the successfully decoded RS codewords as ``side
information"~(prior known information) for the inner decoder. For
$\ell > 1$, such known information can be fed back to the inner
decoder to implement iterations between the outer decoder and the
inner decoder~\cite{Paaske90}\cite{Collins93}. Let ${\rm
Pr}^{(j)}\{\hat{\underline u} \neq \underline u\}$ be the block
error probability after the $j$-th iteration. Obviously, we have
${\rm Pr}^{(j+1)}\{\hat{\underline u} \neq \underline u\} \leq
{\rm Pr}^{(j)}\{\hat{\underline u} \neq \underline u\}$ since such
iterations never change a successfully decoded codeword from a
correct one to an erroneous one.

The decoding algorithm for the RS-Kite codes with incremental
redundancy is described in the following as well as shown in
Fig.~\ref{RSPRLDPCDecoding}.

\begin{algorithm}{(Decoding algorithm for RS-Kite codes)}\label{RSKiteDecAlg}
\begin{enumerate}
    \item {\em Initialization:} Properly choose two positive integers $r_0$ and
    $\Delta_r$. Set $n = k + r_0$. Set $\gamma_i = 0$ for $0\leq i <
    \ell$.

    \item {\em Iteration:} While $n \leq k+T$, do the following
    procedures iteratively.

    {\em \underline {Step 2.1}}: Once the noisy prefix ${\underline y}[n]$ is available in the receiver, perform
    the iterative sum-product decoding algorithm for the prefix Kite code $\mathcal{K}[n,k]$ until
    the inner decoder is successful or the iteration number exceeds a preset maximum iteration
    number $J$. In this step, all successfully decoded RS
    codewords~(indicated by $\gamma_i = 1$) in previous iterations are treated as known information;

    {\em \underline {Step 2.2}}: Set $\delta = 1$. For $0\leq i < \ell$, perform the BM decoding algorithm
    for the $i$-th noisy RS codeword that has not been decoded successfully~(indicated by $\gamma_i = 0$)
    in previous iterations. If the BM decoding is successful, set $\gamma_i = 1$ and $\delta = 0$;

    {\em \underline {Step 2.3}}: If all RS codewords are successfully decoded (i.e., $\gamma_i = 1$ for all $0\leq i < \ell$), stop decoding;
    else if $\delta = 1$, increase $n$ by $\Delta_r$; else keep $n$ unchanged.
\end{enumerate}
\end{algorithm}

{\bf Remark.} In the above decoding algorithm for RS-Kite codes,
the boolean variable $\delta$ is introduced to indicate whether or
not increasing the redundancy is necessary to recover the data
sequence reliably. If there are no new successfully decoded RS
codewords in the current iteration, increase the redundancy and try
the inner decoding again. If there are new successfully decoded
codewords in the current iteration, keep the redundancy unchanged
and try the inner decoding again by treating all the successfully
decoded codewords as known prior information.

\section{Performance Evaluation of RS-Kite Codes and Construction Examples}
\label{sec6}

In this section, we use a semi-analytic method to evaluate the
performance of the proposed RS-Kite code. Firstly, we assume that
the performance of the Kite codes can be reliably estimated by
{\em Monte Carlo} simulation around $\textrm{BER} = 10^{-4}$;
secondly, we use some known performance bounds for RS codes to
determine the asymptotic performance of the whole system.

\subsection{Performance Evaluation of RS Codes}
The function of the outer RS code is to remove the residual errors
in the systematic bits of the inner code after the inner decoding.
Hence, we only consider the systematic bits of the inner code. Let
$P_b$ be the probability of a given systematic bit at time $t$
being erroneous after the inner decoding, that is, $P_b
\stackrel{\Delta}{=} {\rm Pr} \{{\hat v}_t \neq v_t\}$. By
the symmetric construction of Kite codes, we can see that $P_b$ is independent
of the time index $t$~($0 \leq t \leq L_v - 1$). We assume that
$P_b$ can be estimated reliably either by performance bounds or by
{\em Monte Carlo} simulations for low SNRs. To apply some known
bounds to the outer code, we make the following assumptions on the
error patterns after the inner decoding.

\begin{itemize}

    \item The outer decoder sees a memoryless
    ``channel". This assumption is reasonable for large $\ell$.
    For small $\ell$, the errors in the whole inner codeword must be
    dependent. The dependency becomes weaker when only systematic
    bits are considered.

    \item The outer decoder sees a $q$-ary
    symmetric ``channel". This assumption can be realized by taking
    a random generalized RS code~\cite{Moon05} as the outer code
    instead. That is, every coded symbol from the outer encoder is
    multiplied by a random factor $\alpha_i \in \mathbb{F}_q-\{0\}$ prior to entering the inner
    encoder. Accordingly, every (possibly noisy) decoded symbol from
    the inner decoder is multiplied by the corresponding factor $\alpha_i^{-1}$ prior to entering the outer
    decoder.
\end{itemize}

The ``channel" for the outer encder/decoder is now modelled as a
memoryless $q$-ary symmetric channel~(QSC) characterized by $Y = X
+ E$, where $X \in \mathbb{F}_q$ and $Y\in \mathbb{F}_q$ are
transmitted and received symbols, respectively. Their difference
$E \in \mathbb{F}_q$ is a random variable with probability mass
function as
\begin{equation}\label{pmf0}
    P_c \stackrel{\Delta}{=}{\rm Pr}\{E=0\} = (1-P_b)^m
\end{equation}
and
\begin{equation}\label{pmfalpha}
    P_e \stackrel{\Delta}{=}{\rm Pr}\{E=\alpha \} =
    \frac{1-(1-P_b)^m}{q-1}
\end{equation}
for $\alpha \in \mathbb{F}_q - \{0\}$. The decoding error
probability $P_{err}$ of the BM algorithm is given by
\begin{equation}\label{ProbE}
    P_{err} = \sum\limits_{t=t_{max}+1}^{N}
    \left(\begin{array}{c}
      N \\
      t \\
    \end{array}\right) (q-1)^{t}P_e^t P_c^{N-t}.
\end{equation}

For the coding system with $\ell > 1$ RS codewords, successfully
decoded RS codewords will be fed back as prior known information
to the inner decoder. Hence, we need to evaluate the miscorrection
probability of the outer code.  Let $\{A_d, 0\leq d \leq N\}$ be
the weight distribution of the outer code. It is well-known
that~\cite{Moon05}

\begin{equation}\label{weightdistribution}
    A_d = \left(\begin{array}{c}
      N \\
      d \\
    \end{array} \right)
    (q-1)\sum\limits_{i=0}^{d-d_{min}}(-1)^i
    \left(\begin{array}{c}
      d-1 \\
      i \\
    \end{array}\right)
    q^{d-d_{min}-i},
\end{equation}
where $d_{min} = N-K+1$. The miscorrection probability can be
calculated as~\cite{Sofair00}
\begin{equation}\label{probmis}
    P_{mis} = \sum
    A_d\left(\begin{array}{c}
      d \\
      i \\
    \end{array}\right)
    \left(\begin{array}{c}
      d-i \\
      j \\
    \end{array}\right)
    \left(\begin{array}{c}
      N-d \\
      h \\
    \end{array}\right)
    (q-2)^j(q-1)^h P_e^{i+j+h}P_c^{N-i-j-h},
\end{equation}
where the summation is for all $d, i, j, h \geq 0$ such that
$d_{min} \leq d \leq N$, $i\leq d$, $j \leq d-i$, $h\leq N-d$,
$i+j+h > t_{max}$ and $d-i+h \leq t_{max}$. Under the worst case
assumption that $P_b = 0.5$, the miscorrection probability is
upper bounded by $\frac{1}{t_{max}!}$~\cite{McEliece86}.

\begin{figure}
\centering
\includegraphics[width=10cm]{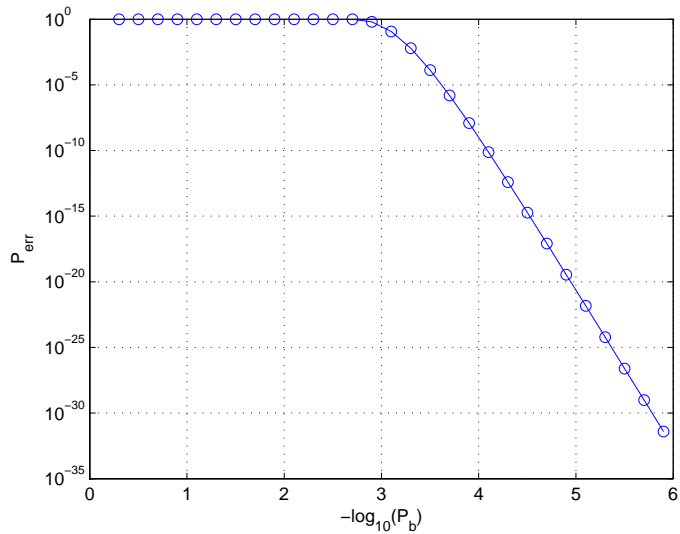}
\caption{Probability of decoding error for the RS code
$\mathcal{C}_{1024}[1023, 1000]$ using bounded distance decoding.}
\label{Pdec10231000}
\end{figure}

\begin{figure}
\centering
\includegraphics[width=10cm]{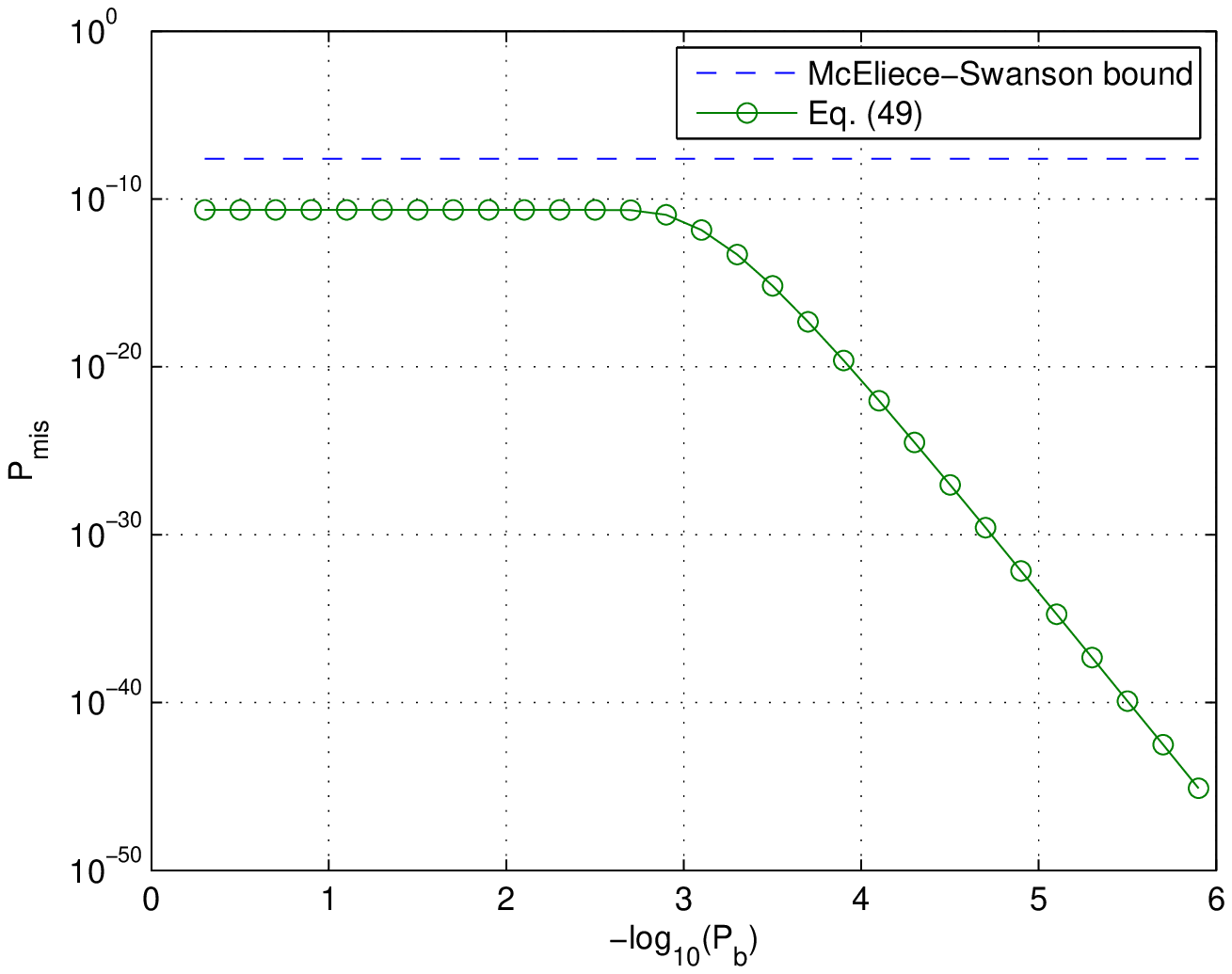}
\caption{Probability of miscorrection for the RS code
$\mathcal{C}_{1024}[1023, 1000]$ using bounded distance decoding.}
\label{Pmis10231000}
\end{figure}

\begin{figure}
\centering
\includegraphics[width=10cm]{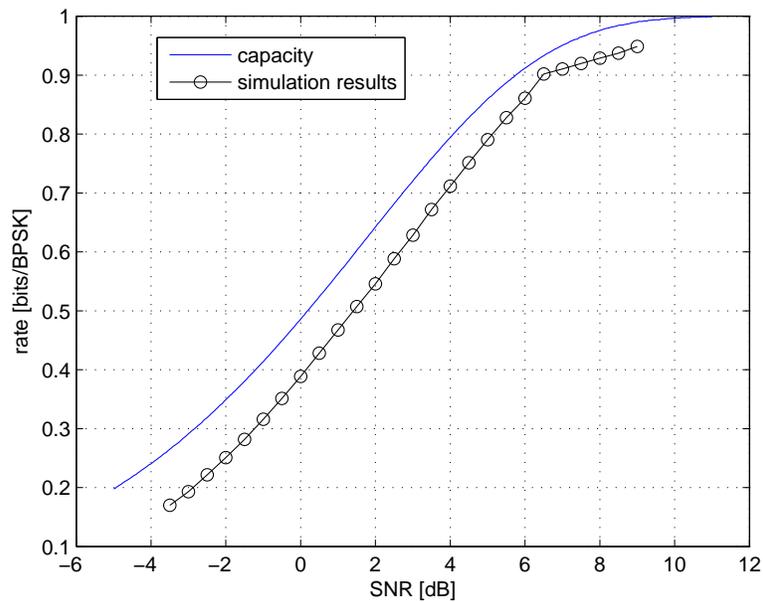}
\caption{The average decoding rates of the Kite codes. The data
length $k = 50000$.} \label{rateless}
\end{figure}

\subsection{Construction Examples}

We take RS code $\mathcal{C}_{1024}[1023, 1000]$ over
$\mathbb{F}_{1024}$ as the outer code. This code can correct up to
$11$ symbol-errors. The performance of this code is shown in
Fig.~\ref{Pdec10231000} and Fig.~\ref{Pmis10231000}, respectively.
Also shown in Fig.~\ref{Pmis10231000} is the McEliece-Swanson
bound~\cite{McEliece86}. This computations can be used to predict the performance of the RS-Kite codes.
For example, if the inner Kite code has an error-floor at $\textrm{BER} = 10^{-4}$, then
the outer RS code $\mathcal{C}_{1024}[1023, 1000]$ can lower down this error-floor to $10^{-10}$ with
miscorrection probability not greater than $10^{-20}$. This prediction is very reliable for large $\ell$.

We set $\ell = 5$. Then the length of the
input sequence to the inner code is 51150. We take Kite code
$\mathcal{K}[\infty, 51150; \underline p]$ as the inner code. The
$p$-sequence is specified by
\begin{equation}\label{pSequence51150}
    p_t = \left\{\begin{array}{rl}
      q_9 = 0.00084, & 0.9 \leq k/(t+k)<  1.0  \\
      q_8 = 0.00020, & 0.8 \leq k/(t+k) <  0.9 \\
      q_7 = 0.00015, & 0.7 \leq k/(t+k) <  0.8 \\
      q_6 = 0.00009, & 0.6 \leq k/(t+k) <  0.7 \\
      q_5 = 0.00006, & 0.5 \leq k/(t+k) <  0.6 \\
      q_4 = 0.00006, & 0.4 \leq k/(t+k) <  0.5 \\
      q_3 = 0.00004, & 0.3 \leq k/(t+k) <  0.4 \\
      q_2 = 0.00002, & 0.2 \leq k/(t+k) <  0.3 \\
      q_1 = 0.00001, & 0.1 \leq k/(t+k) <  0.2 \\
    \end{array}\right..
\end{equation}

The simulation results are shown in Fig.~\ref{rateless}, where the
error probability for each simulated point is upper-bounded by
$P_{mis}$ because the receivers will try decoding with increasing
redundancies until all RS codewords are decoded successfully.
In our simulations, we have not observed any decoding errors. It
can be seen from Fig.~\ref{rateless} that the gaps between the
average decoding rates and the capacities are around $0.1$
bits/BPSK in the SNR range of $-3.0\sim 9.0$ dB.
To our best knowledge, no simulation results were reported in the literature
to illustrate that one coding method can produce good codes in such a wide range.
We have also observed
that there is a ``singular" point at rate of $0.9$. This is because we have
taken $0.9$ as the first target rate. If we add one more target rate of $0.95$,
we can make this curve more smooth. But the other $q$-parameters need to be re-selected
and the whole curve will change.

\section{Conclusion}\label{conclusion}
In this paper, we have proposed a new class of rateless forward
error correction codes which can be applied to AWGN channels. The
codes consist of RS codes and Kite codes
linked in a serial concatenation manner. The inner Kite codes
can generate potentially infinite parity-check bits with linear
complexity. The use of RS codes as outer codes not only lowers
down the error-floors but also ensures with high probability the
correctness of the successfully decoded codewords. A semi-analytic method has
been proposed to predict the error-floors of the proposed codes,
which has been verified by numerical results. Numerical results
also show that the proposed codes perform well over AWGN
channels in a wide range of SNRs in terms of the gap to the
channel capacity.


%

\appendices


\section*{Acknowledgment}
The authors would like to thank X. Huang, J.-Y. You and J.~Liu for their help.
%

\ifCLASSOPTIONcaptionsoff

\newpage
\fi



%

\bibliographystyle{ieeetr}
\bibliography{tzzt}




%

%
%
%




\end{document}